# Visualisation of massive data from scholarly Article and Journal Database: A Novel Scheme

Gouri Ginde[1]

Department of Computer Science and Engineering,
PESIT Bangalore South Campus, Bangalore, India [1]

**Abstract**
Scholarly articles publishing and getting cited has become a way of life for academicians. These scholarly publications shape up the career growth of not only the authors but also of the country, continent and the technological domains. Author affiliations, country and other information of an author coupled with data analytics can provide useful and insightful results. However, massive and complete data is required to perform this research. Google scholar which is a comprehensive and free repository of scholarly articles has been used as a data source for this purpose. Data scraped from Google scholar when stored as a graph and visualized in the form of nodes and relationships, can offer discerning and concealed information. Such as, evident domain shift of an author, various research domains spread for an author, prediction of emerging domain and sub domains, detection of journal and author level citation cartel behaviours etc. The data from graph database is also used in computation of scholastic indicators for the journals. Eventually, econometric model, named Cobb Douglas model is used to compute the journal's Modeling Internationality Index based on these scholastic indicators.

**Keywords**: Data acquisition methods, Web scraping, Graph database, Neo4j, Data visualization, Cobb Douglas model.

## I. INTRODUCTION

Data visualisation is the visual representation of patterns in data. It can help to understand and relate to the data, communicate and represent data in a more comprehensive manner to others. Data visualisation can be as trivial as a simple table, elaborate as a map of geographic data depicting an additional layer in Google Earth or complex as a representation of Facebook's social relationships data. Visualisation can be applied to qualitative as well as quantitative data. Visualisation has turned into an inexorably well-known methodology as the volume and complexity of information available to research scholars has increased. Also, the visual forms of representation have become more credible in scholarly communication. As a result, increasingly more tools are available to support data visualisation.

### A. Need for effective visualization of scholarly publications

Scholarly articles and journals have always been a subject of interesting research, mainly because of the stakes involved in the scholarly publications primarily in the world of academia.

These articles are weighed based on the credibility of the journals first and then the authors. Nevertheless there does not exists one complete solution which can provide complete visualization of the data at author, country, and journal and domain level for all the journals in the world.
This data, when explored using right tools, has a tremendous potential to unearth interesting patterns which can expose illegitimate cartel behaviour at various levels as well as enunciate out performing under mined authors and evolving journals.
Massive scale journals data can be used for the purpose of a data visualisation is analysis, communication, or both. Analysis requires careful attention to the parameters used. Different parameters reveal different patterns, and it is challenging to determine which are significant with respect to the key research questions.

### B. Importance of data visualization

High impact visualization is like a picture speaking a thousand words. Selecting good visual technique to display the data holds key to a good impact. Fancy bubble charts, Time domain based motion graphs are possible now because of the languages such as python and R. However, selecting the one which can effectively represent the data to the audience is crucial. To begin with, in this paper we have visualized data using in-built D3.js script available with Neo4j Graph database. Further on we have used the pie chart, line graph and area graphs to represent the information concisely to overcome data overload through dynamic presentation.in the initial stage.
Fig 1 shows the flow diagram of the complete system. A huge data from Google scholar has been scraped using web scraping methodology to acquire the required dataset over a period of 2 Months (cite our paper). Further this data is pre-processed and fed into Neo4j graph database using advanced cyphers. These cyphers deconstruct complex JSON documents and quickly turn them into a graph structure of rich relationships without duplication of information. In the next stage the data from the graph database is queried using various Cyphers to compute a few more scholastic indicators such as self-citations, total citations, international collaboration ratio etc.

These scholastic indicators are then created as the properties at article and journal level in the graph database. Finally, various questions are transformed into Cypher queries to get the meaningful data from the graph database. This data is then visualized using various charts, graphs etc.

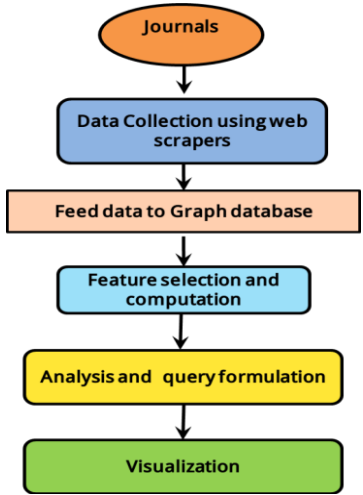

Fig 1. Flow diagram

## II. DATA CURING

Data accumulation, a first step in data curing, is an arduous task for any research project. For this research Google scholar has been used as resource as it provides a comprehensible and complete data required for good analysis. However, Google scholar does not provide any API. Hence web scraping methodology has been used to gather the data [1, 4]. Next, accumulated data is pre-processed, an intermediate task where the data is cured and made ready for further analysis. The scraped data, which is in JSON format, is first trimmed of any unwanted characters. Then, data is cleansed of Unicode characters.

In order to provide interesting visualization patterns, few of the parameters had to be derived from accumulated data. Pre-processing also involves computation of these scholastic indicators/parameters. The scraped data, which is in JSON format, is in the pure textual form hence, cosine similarity string metric has been used for text comparison in place of pure string comparison operation for better results.

Using cosine similarity string metric, Author-level and journal-level Self citation, International Collaboration ratio and other scholastic indicators are computed. Self-citation count is a part of citation count when a citing article shares at least one author name with the article it is citing. journal-level self-citation is part of citation count when a citing article shares the Publishing journal name same as that of an article it is citing.

## III. COBB DOUGLAS MODEL FOR COMPUTATION OF 'INTERNATIONALITY'

**Definition of 'Internationality' of a journal**: Internationality has been defined and perceived as the degree to which a journal transcends local communities and boundaries, with respect to the quality of publication and influence. We define internationality of peer-reviewed journals as a measure of influence that spreads across boundaries and attempts to capture different and hitherto unperceived aspects of a journal for computing internationality. Internationality, $y$ is defined as a multivariate function of $x_i, i=1, 2...n$. Internationality score varies over time and depends on scholastic parameters, subject to evaluations, constant scrutiny and ever changing patterns.

**Cobb Douglas Model:** In economics, Cobb-Douglas production function [2,5,6] is widely used to represent relationship of outputs to inputs. This is a technical relation which describes the Laws of Proportion, i.e., the transformation of factor inputs into outputs at any particular time period. This production function is used for the first time, to compute the internationality [3] of a journal where the predictor/independent variables, $x_i, i=1, 2...n$ are algorithmically extracted from curated data.

Cobb-Douglas function is given by:

$$y = A \prod_{i=1}^{n} x_i^{\alpha_i}$$

Where, $y$ is the internationality score, $x_i$ are the predictor variables/input parameters and $\alpha_i$ are the elasticity coefficients.

The function has extremely useful properties such as convexity/concavity depending upon the elasticity's. The properties yield global extrema which are intended to be exploited in the computation of internationality or influence. For $n = 4$, $x1$ to $x4$ are the input parameters as described below.
- $x1$: other-citations quotient)
  = [1 - (self-citations /Total citations)]
- $x2$ : International Collaboration
- $x3$ : Source Normalized Impact per paper (SNIP)
- $x4$ : Non-Local Influence Quotient
  = [1 - (Journal's self-citations /Total citations)]

## IV. GRAPH DATA MODELING

Data accumulated is rich, very well connected and has a lot of hidden information within it. Hence, we chose to visualize this data using graph database. A graph database is a graph-oriented database, which is type of NoSQL database that uses graph theory to store, map and query relationships. It is basically a collection of nodes and edges. [1,7]

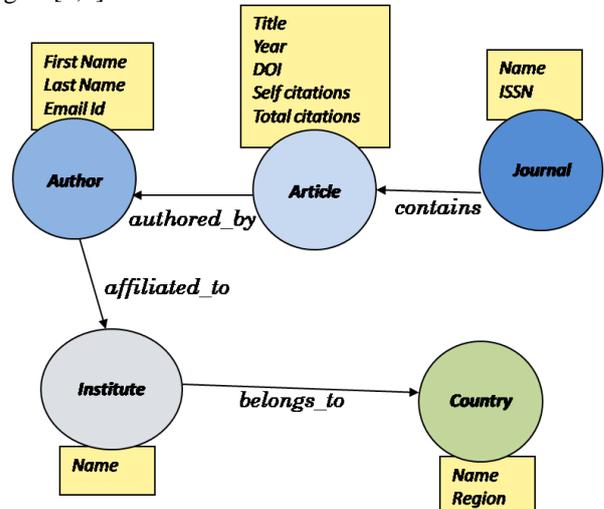

Fig 2 Data model for Graph database

Graph data modeling [8] is the process in which a Neo4j user describes an arbitrary domain as a connected graph of nodes and relationships. From this description, a graph data model is designed to answer questions in the form of Cypher queries.

Scholarly articles and scientific journals make the most of our research area hence we identified the elements from these which can then be transformed into nodes and relationships. Few of the elements constitute properties as shown in the square boxes beside the oval shaped nodes. Fig 2 is the data model that has been designed using the Neo4j graph data modeling.

## V. DATA VISUALIZATION

Scraped data is imported into graph database of Neo4j. Few of the visualizations that are possible with our data model are:
- Author network
- Institute network
- Country network
- Spread of a domain in a country
- Collaboration network of an institute
- Extract year-wise publication trend of an author

This data when queried appropriately can help to visualize the shape of 'Internationality' of a Journal at various levels. Few of the visualization are as following.

**Journal to Author to Country mapping:** Figure 3 shows mapping of a journal to the Author and in turn to the Country to which his/her Institution belongs to.

The blue circles represent journal nodes, purple circles are author nodes, yellow represent article nodes and red ones are country nodes. These links will actually help in identifying how well countries of a region are contributing to a domain. Since we can identify spurious journals using journal's internationality modeling index (JIMI), we can now identify which regions are actually contributing more to such nexus of fake and dubious journals.

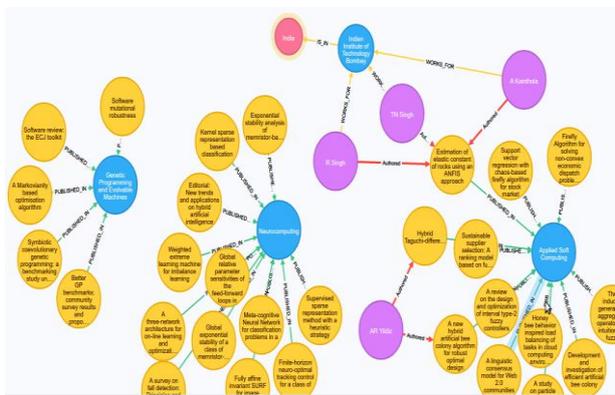

Figure 3. Journal to Author to Country mapping

**Author to Institute to Country mapping:** Figure 4 shows mapping of the Author to Institute to Country. Blue nodes represent Institution, red represent country and purple represent author nodes. This mapping can help in identifying the contribution trends pertaining to a particular Institute and Country in particular.

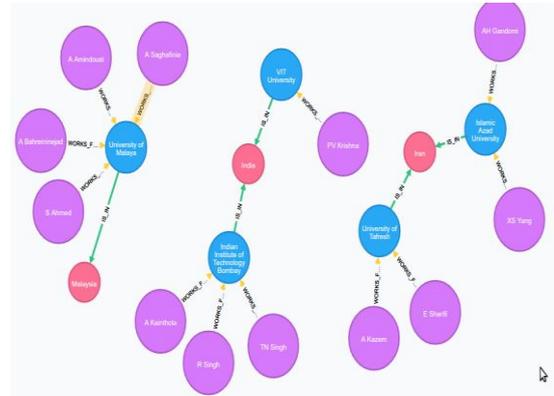

Figure 4. Author to Institute to Country mapping

**Article to Author Mapping:** Figure 5 is visualization of a particular author's contribution in totality. We can query this data based on year to visualize year by year contributions made. In the figure purple nodes are authors and yellow nodes are articles. For a rich and dense data we can provide visually appealing information about an author's reach and contributions using this visualization.

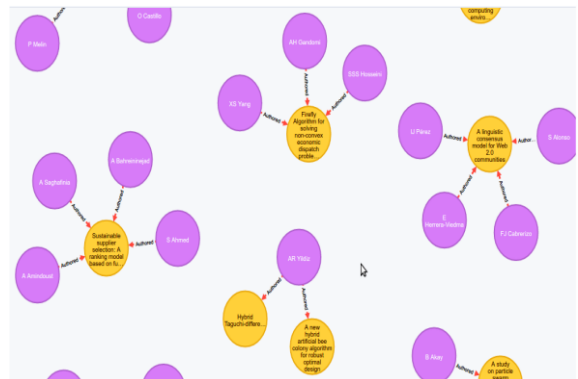

Figure 5 Author to Article mapping

**Institute to Country to Region mapping:** Figure 6 is the data for all of the institutes belonging to a country, and countries belonging to a Region. These links can help in identifying how well countries of a region are contributing to a domain. Since we can identify spurious Journals, we can now identify which regions are contributing more to such nexus of fake and dubious journals. All the articles with a degree of Article impact, when measured using Singular value Decomposition for that domain and the journals with a measure of 'internationality' (computed using Cobb Douglas model) will define the degree of contribution made by any Institute and in turn Country to a particular scientific domain (to which that journal belongs to).

In other words, these two measures i.e. Article impact and the journal's 'internationality' index can be used to define contribution of any Author, Institute, Country and Region to a particular domain.

Conversely these two parameters can be used as a scale to evaluate Authors, journals, Institutes, Countries and Regions. This scale can explain the growth and contribution made by all of these. When the data is large enough we can predict evolving field, most dominating country in a particular field/domain and increase or decrease of impact for any given journal.

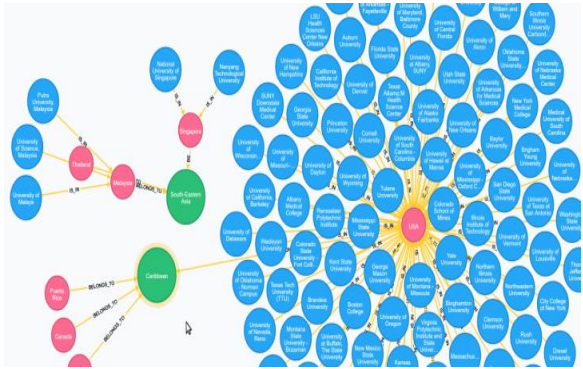

Figure 6. Institute to Country to Region mapping

## VI. RESULTS

Following are the various cypher queries and resulting visualization from the graph database. Fig 7 depicts journal to Author and in turn to the Country of affiliation mapping. Following is the query on the graph database to extract the needed information for graph plotting.

**Cypher Query:**
MATCH (Journal)<-[:PUBLISHED_IN]-(Article) WHERE Journal.name IN ['Applied Soft Computing', 'Neurocomputing', 'Genetic Programming and Evolvable Machines'] RETURN Article.year, Journal.name

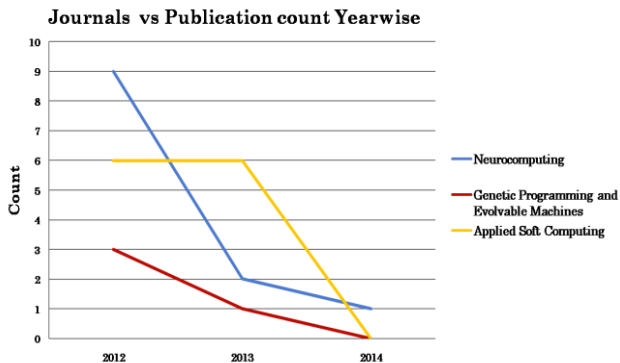

Figure 7. Line graph; Journal Vs Publication

Fig 8 shows the area graph of total citations vs. self-citations for all the articles and journals in the graph database. Following is the query which extracts this data.

**Cypher query:**
MATCH (n:Article) RETURN n.totalcites, n.selfcites

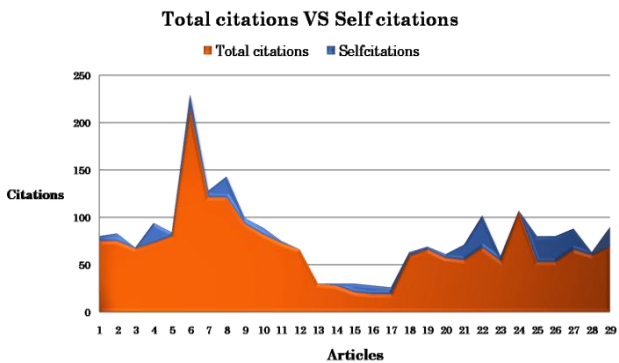

Figure 8. Area graph; Total citation Vs. Self citations

Fig 9 shows the Pi graph of Article Publications per Country. Following is the query which is executed on the database to extract this data.

**Cypher Query:**
MATCH (Author)-[r:WORKS_FOR]->(Institute)-[s:IS_IN]->(Country) RETURN Author.name, Country.name

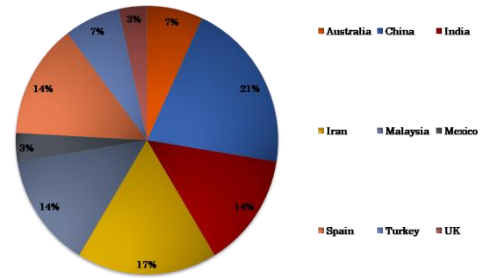

Figure 9. Pie graph; Article Publications per Country

## VII. CONCLUSION

Scholarly articles and Scientific journal datasets need special type of database as the data is massive in scale and ever evolving. Various web-scraping and parsing techniques were used to create and develop a platform for ScientoBASE [7], a repository, which will consist of international journals by subject category with ranks and scores of internationality and necessary metric information. Graph database such as Neo4j, which is a NoSQL type of database is an emerging technology in the field of effective visualization and data storage. It not only provides a more meaningful method of data storage but also facilitates intelligent query formulation for the meaningful data extraction and analysis. In this research Neo4j has played a crucial role in finding the hidden patterns which can further enhance the usability of the information concealed within the huge databases maintained worldwide. This work will lead to software which will be an end-to-end product comparable with Scopus and ISI's Web of Science but positioned in a distinct space and cater to the needs of the underprivileged researchers in developing countries.

The broader aim of our research is to define a yardstick of scientific contribution and international diffusion; especially in niche areas such as Astroinformatics, Computational Neuroscience, Industrial Mathematics and Data Science from India, as well as other countries across the globe. The outcome of our research will pave way for data and model validation and construction of a data visualization and web interface tool (ScientoBASE Toolkit), an open source web interface that will compute the scores and provide visualizations of all essential parameters of internationality. This tool can be used as a web-kit to measure/analyze the growth of Indian as well as global Scientometry in state of the art and emerging areas in Science and Technology.